\def\arcdeg{\hbox{$^\circ$} }

\newcommand{\micron}{\,\hbox{$\mu$m} }
\newcommand{\micronn}{\,\hbox{$\mu$m}}
\def\etal{ {\em et~al.\/}\thinspace}

\def\kms{ km~s$^{-1}$}

\documentstyle[11pt,aaspp4,epsfig,flushrt]{article}
\lefthead{Monnier et al.}
\righthead{RF Filterbank for Mid-IR Interferometry}

\begin{document}

%_____________________________________TITLE PAGE___________________________
\title{Mid-infrared interferometry on spectral lines:
\\ I. Instrumentation}
\author{J. D. Monnier\altaffilmark{1}, W. Fitelson,  
W. C. Danchi\altaffilmark{2} and C. H. Townes}

\altaffiltext{1}{Current Address: Smithsonian Astrophysical Observatory 
MS\#42, 60 Garden Street, Cambridge, MA, 02138}

\altaffiltext{2}{Current Address: NASA Goddard Space Flight Center,
Infrared Astrophysics, Code 685, Greenbelt, MD 20771}

\affil{Space Sciences Laboratory, University of California, Berkeley,
Berkeley,  CA  94720-7450}

%
%
%
%____________________________________ABSTRACT PAGE_________________________
\begin{abstract}

The U. C. Berkeley
Infrared Spatial Interferometer has been outfitted with 
a filterbank system to allow interferometric observations of mid-infrared
spectral lines with very high spectral resolution
($\frac{\lambda}{\Delta\lambda}\sim10^5$).  This paper describes the
design, implementation, and performance of the matched 32-channel
filterbank modules, and new spectral line observations of Mars and
IRC\,+10216 are used to demonstrate their scientific capability.  In
addition, observing strategies are discussed for accurate calibration
of fringe visibilities in spectral lines, despite strong atmospheric
fluctuations encountered in the infrared.  
\end{abstract}


%_______________________________________INTRODUCTION_______________________
%\pagebreak
\section{Introduction}

The Infrared Spatial Interferometer (ISI) consists of two
1.65-m telescopes employing heterodyne detection between
9--12\,$\mu$m using CO$_2$
lasers as local oscillators (LO) (\cite{hale2000};
\cite{lipman98}).  Observations can be made at a large number of
frequency bands in this spectral range by tuning to the various
lasing transitions of different CO$_2$ isotopes.  The ISI has measured
the characteristics (sizes and optical depths) of dust shells around
late-type stars (e.g., \cite{danchi94}) and angular diameters of the
nearest red supergiants (\cite{bester96}).  In these cases,
sensitivity to continuum (dust or photospheric) emission using
heterodyne detection is generally less than that theoretically
possible using direct (photon-counting)
detection techniques, partly because the maximum
bandpass is limited by the temporal response of the IR detector,
which is about 5\,GHz with present detectors
($\frac{\lambda}{\Delta\lambda}\sim5000$).

However for narrow bandpasses, heterodyne receivers are
more sensitive than direct detection schemes, which are not
background-limited for low
flux levels (see more detailed discussion by Monnier [1999] and
Hale\etal [2000]).  Furthermore, extremely high spectral resolution is
attainable with heterodyne spectroscopy (e.g., Betz 1977) since the
down-converted astronomical signal can be further filtered with
conventional radio-frequency (RF) filters.  
Such high resolution is needed to resolve
many molecular lines which form in cool winds around AGB stars 
($\Delta$v$\sim$1\kms) or in planetary atmospheres.  
Betz was able to observe transitions of CO$_2$ around Mars and Venus
with spectral resolution of $\frac{\lambda}{\Delta\lambda}\sim$6$\times$10$^6$
(\cite{betz76}).
Interferometry with this resolution can hence measure the location
and distribution of molecules in particular excitation states.

Naturally, the high spectral resolution potential of a heterodyne
system can only be realized by using a spectrometer
in the RF signal chain following detection, a capability not part of the 
original ISI design.
Although previous systems have been built for high
efficiency spectral line work with wide RF bandwidths on single
telescopes (e.g., \cite{betz77}; \cite{goldhaber88}; \cite{isaak99};
\cite{holler99}), no system existed that was suitable for 
interferometry on narrow bands because of the need to correlate
signals from two separate telescopes.

This paper will describe the implementation of an inexpensive
filterbank system for spectral line observations using the ISI.  After
a description of the filterbank itself, 
its performance will be
evaluated from laboratory measurements and test observations of
astronomical objects.  
Fluctuations from the turbulent
atmosphere and unexpected
instrumental drifts required new observing strategies for accurate
calibration of the fringe visibility on and off of spectral lines.
These techniques are described in detail, including fast bandpass
switching for phase-referencing and LO-switching for robust correlator
calibration.  Lastly, the near-term scientific potential of combining
high spectral and spatial resolution at these wavelengths is reviewed.

\section{The 32-Channel Filterbank}
\subsection{Introduction}
In principle, heterodyne spectrosocopy with an RF filterbank is quite simple, even 
at 27~THz.  A CO$_2$ laser in each ISI telescope acts as a local oscillator, with
a wavelength of $\sim$11\,$\mu$m.  Wavefronts from the laser are mixed with light from
the sky and the resulting ``beat'' pattern is 
detected by a HgCdTe photoconductor with a large output bandwidth
($\pm \Delta\nu$).  This process is called ``heterodyne'' detection and
down-converts a $2 \Delta\nu$ bandwidth centered
around 27~THz (mid-infrared radiation)
into microwave signals between DC and $\Delta\nu$.  In a double-sideband (DSB) system 
such as the ISI, frequencies both above and below the local oscillator frequency
are converted into the same (microwave) frequency, thus 
overlapping each other.
When observing a spectral line, this has the undesired 
effect of diluting the line (in one
sideband) with uninteresting continuum radiation (from the other sideband), but 
can still be readily interpreted when multiple spectral features are not overlapping.

Once the signals have been down-converted to microwave (or RF) frequencies, they
are amplified by cold FET amplifiers and transmitted through coaxial cables.  
Tunable RF filters with bandwidths of 60\,MHz 
(and much narrower)
have been available for quite some time, and
can be used to filter the astrophysical signals.  Since the original
frequency of the radiation was $\sim$27\,THz, the use of filters with 60\,MHz 
bandwidth corresponds to a spectral resolving power of
 $\frac{\lambda}{\Delta\lambda}=\frac{27\,{\rm THz}}{60\,{\rm MHz}} 
= 4.5\times10^5$, fine enough to resolve spectral features arising from doppler
shifts as small at $\sim$0.7\,km\,s$^{-1}$.  

\subsection{Design Considerations}
Being a first generation instrument,
the design goals of the ISI filterbank were modest:
\begin{itemize}
\item{Spectral resolution must be sufficient to resolve
narrow absorption lines around AGB stars (a few km/s)}
\item{Full bandpass must be broad enough to encompass both the 
relevant absorption line and significant bandwidth of
continuum ($\sim$2\,GHz)}
\item{Due to high demand for ISI continuum observations,
the filterbank system 
should minimally interfere with standard
observing, i.e. should not involve a radically different
hardware configuration}
\item{A fast bandpass switching scheme
to calibrate atmospheric fluctuations should be implemented}
\item{Cost must be relatively low, -- i.e., no digital correlator}
\end{itemize}

\begin{figure}
\begin{center}
\centerline{\epsfxsize=5in{\epsfbox{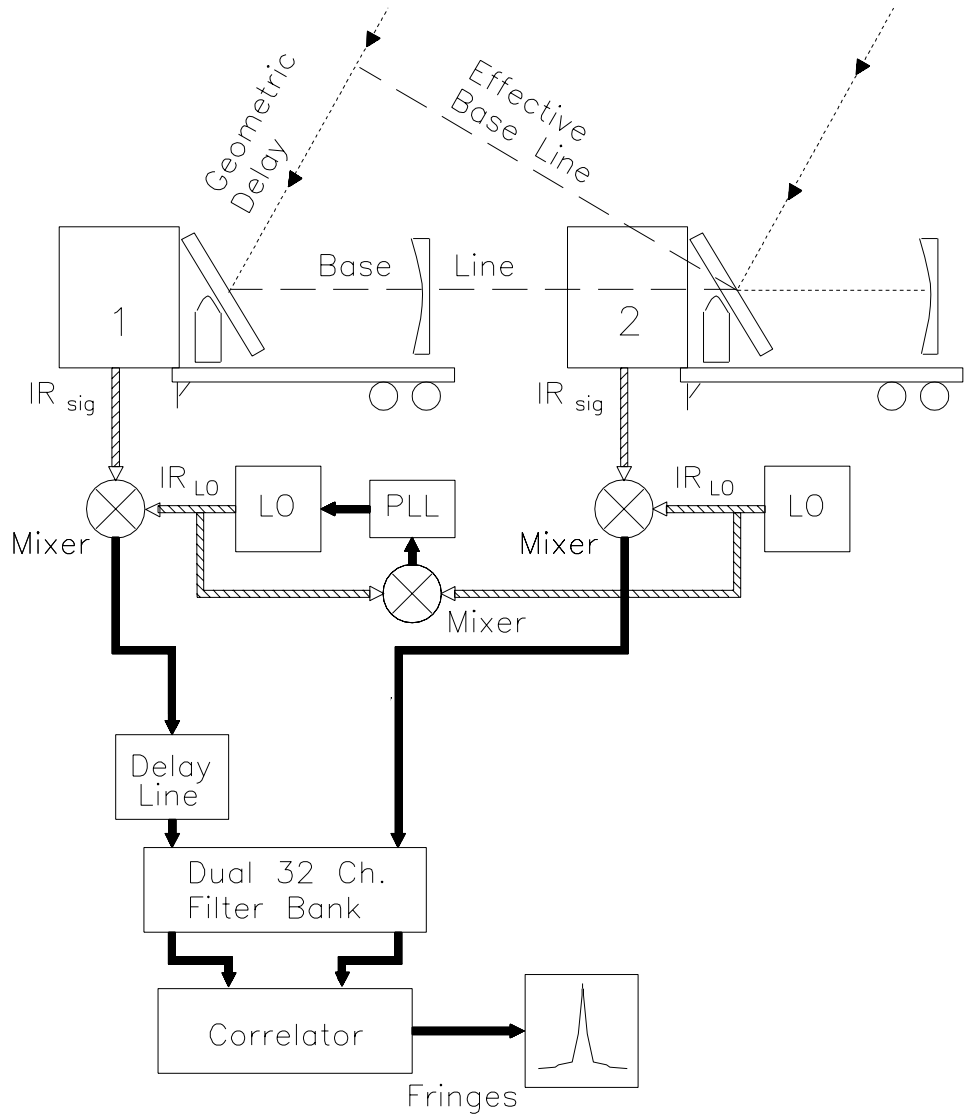}}}
\caption[Block Diagram of ISI]{This figure shows a block
diagram of the ISI interferometer with the filterbank in the RF path.
Two trailers, each containing a telescope are indicated.  Infrared
(IR) signals are mixed with local oscillator (LO) power from 
CO$_2$ lasers.  After passing through a filterbank which determines the
bandpass, the resulting radio frequency (RF) signals from the two telescopes 
interfere in a correlator, producing interference fringes.  The two LOs are
kept in phase by a Phase Lock Loop (PLL) circuit.  See Hale\etal (2000) for
complete details.
\label{fig:isi}}
\end{center}
\end{figure}

\begin{table}
\caption[ISI Filterbank specifications]{Important
Specifications of ISI Filterbank
\label{table:FB_chars}}
\begin{center}
\begin{tabular}{cc}
\hline
Bandpass  & 270-2190 MHz \\
Number of Filters       & 32 per telescope \\
Individual Filter Bandwidth & 60 MHz \\
(3dB points) & \\
Spectral Resolution at 11.15\,$\mu$m & $\sim$0.7 km s$^{-1}$ \\
Phase Matching & $\pm$10$\arcdeg$ over 80\% of band \\
               & $\pm$20$\arcdeg$ over entire band \\
Gain Dynamic Range & 15 dB \\
Bandpass Switching Rate &  up to 500 Hz \\

\hline
\end{tabular}
\end{center}
\end{table}

Table\,\ref{table:FB_chars} summarizes the technical features of the
completed filterbank system, satisfying all of the above design goals.  
The desire to minimally impact the current hardware
design of the ISI imposed the most
severe restrictions on the design.  In order
to maintain the current RF architecture behind the correlator and to
avoid the development of an entirely new data acquisition system,
spectroscopic capabilities were added to the ISI by inserting a
filterbank system before the correlator and total infrared power detection
subsystems.  Figure\,\ref{fig:isi} shows a schematic of the major
subsystems of the Infrared Spatial Interferometer (ISI), including the
dual 32-channel filterbank.  Detailed descriptions of the ISI itself 
can be found in the instrument paper by Hale\etal (2000)
and in recent PhD theses (\cite{lipman98}; \cite{mythesis}).

\begin{figure}
\begin{center}
\centerline{\epsfxsize=\columnwidth{\epsfbox{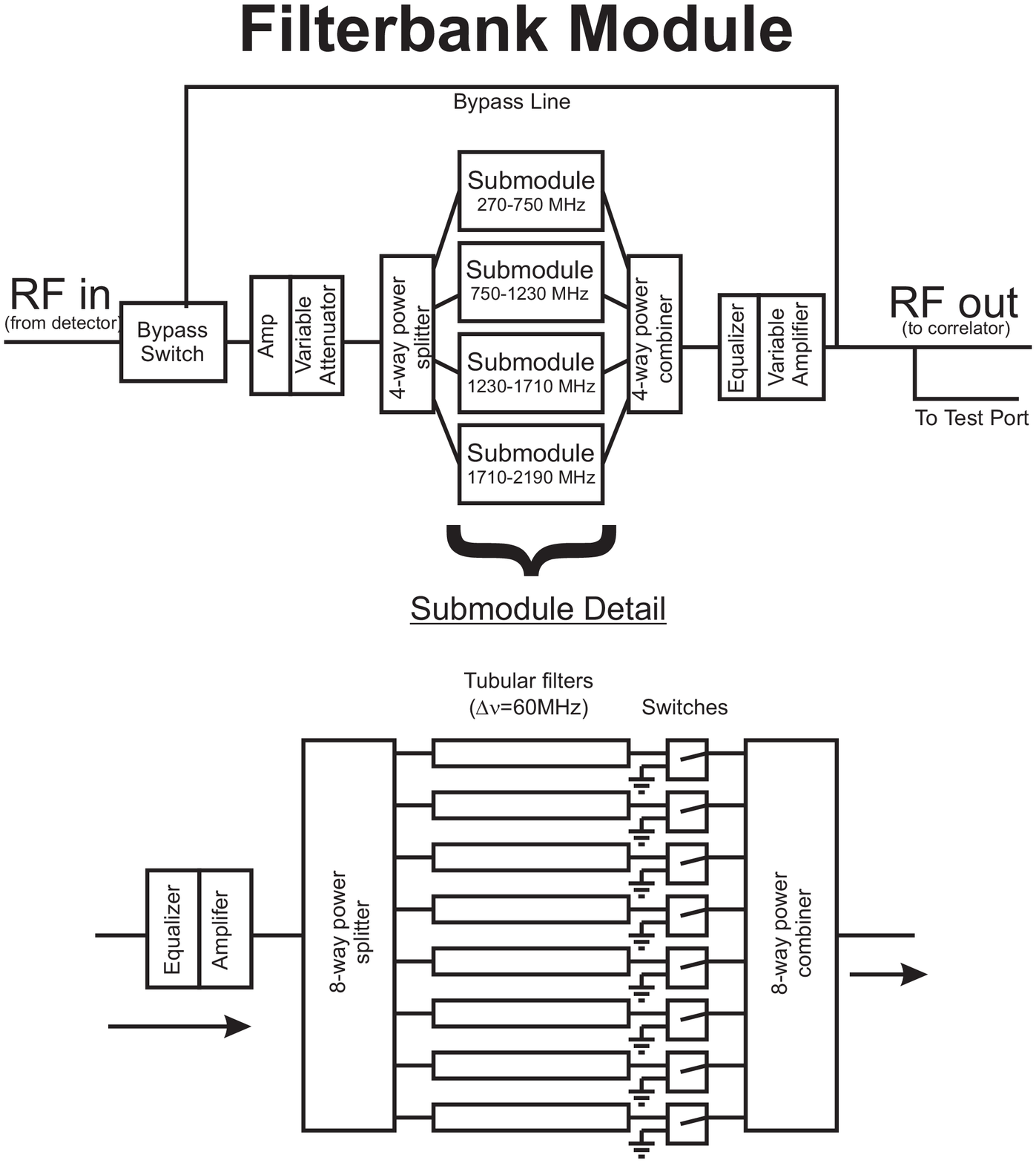}}}
\caption[Detailed block diagram of filterbank]{This figure shows a block
diagram of one of the filterbank modules; there is one per telescope.
\label{fig:FB_detail}}
\end{center}
\end{figure}

The interferometer must be able to measure not only the flux density
of infrared radition incident on both telescopes (i.e., the total IR
power), but also the strength of the interference of signals from both
telescopes (i.e., the fringe power).  By filtering the RF signals from
both telescopes with the two filterbank modules before correlation,
the infrared (IR) power and fringe signal could be measured in a
manner identical to that used for the broadband (continuum) signal.
When not performing spectroscopic observations, the filterbank can be
switched out of the RF path through a bypass line inside each
filterbank module (see figure\,\ref{fig:FB_detail}).  This strategy
has the distinct advantage of not impacting the existing hardware {\em
at all}, and also does not require any additional RF detection or data
acquisition development.  However, because the filterbank intercepts
the RF signals before correlation, the two modules of the filterbank
(one set of 32 filters for each telescope) must be phase-matched.
This imposed strict requirements on all the RF components inside the
filterbank and contributed significantly to construction complexity
and component costs.  The other important performance trade-off was
that only one filter bandpass can be observed at a time, unlike
typical spectrometers where detectors (e.g., RF diodes) are placed
behind all filters and read out in parallel, allowing the entire
spectral line to be measured at once.  This limits the uses of the
filterbank, for example making it impractically slow to perform standard
spectroscopic observations for a large number of lines.

Figure\,\ref{fig:FB_detail} shows a detailed block diagram of one of
the two filterbank modules, with every major component in the RF chain.  
In order to filter the bandpass, the
original RF signal from each telescope is split into 32 frequency
bands via a 4-way power splitter followed by four 8-way splitters.
This decreases the signal power in each band by 15 dB, and
amplification is applied before, during, and after this splitting to
keep signal levels high enough that the additional amplifier noise
from this part of the signal chain causes no significant decrease to
the signal-to-noise ratio (SNR).  In addition, custom frequency
equalizers are used to maintain a reasonably flat frequency response
throughout each module and within each sub-module.  After being split
into 32 bands, the RF signals encounter the filters.  Thirty-two
switches are used to select the desired bandpass (any combination is
allowed), after which the signals are recombined through a symmetric
combination of four 8-way combiners and a single 4-way combiner.
Following recombination, the signal is amplified and any overall
bandpass slope removed before finally being re-injected into the ISI
system.

In order to make sure that the filterbank puts out as much RF power as
it receives (net power gain of 0 dB) independent of the bandpass
selected, a variable attenuator and amplifier are used in combination
to provide up to 15~dB of relative amplification.  This is done so that
the laser shot-noise always dominates the measurement noise, and that the
final detection diodes are always used at similar power levels.  

Onboard memory allows the storage of two separate bandpass selections,
which can be switched back and forth rapidly either by a 
computer-controlled chop signal or one from an external signal generator.
This capability allows the measurement of both the narrow absorption line
and the broad continuum on alternate chop cycles
within an atmospheric coherence time, removing
brightness fluctuations as a source of measurement uncertainty.
Bandpass selection and gain/attenuator control can be accomplished
via manual switching  or through a computer interface.
Presently, the filterbank bandpass and gain selections are made using
a C-language program running on a Sun workstation 
interfaced with the main ISI control computer.  

\subsection{Performance}
\label{section:fb_performance}
\begin{figure}
\begin{center}
\centerline{\epsfxsize=5.1in{\epsfbox{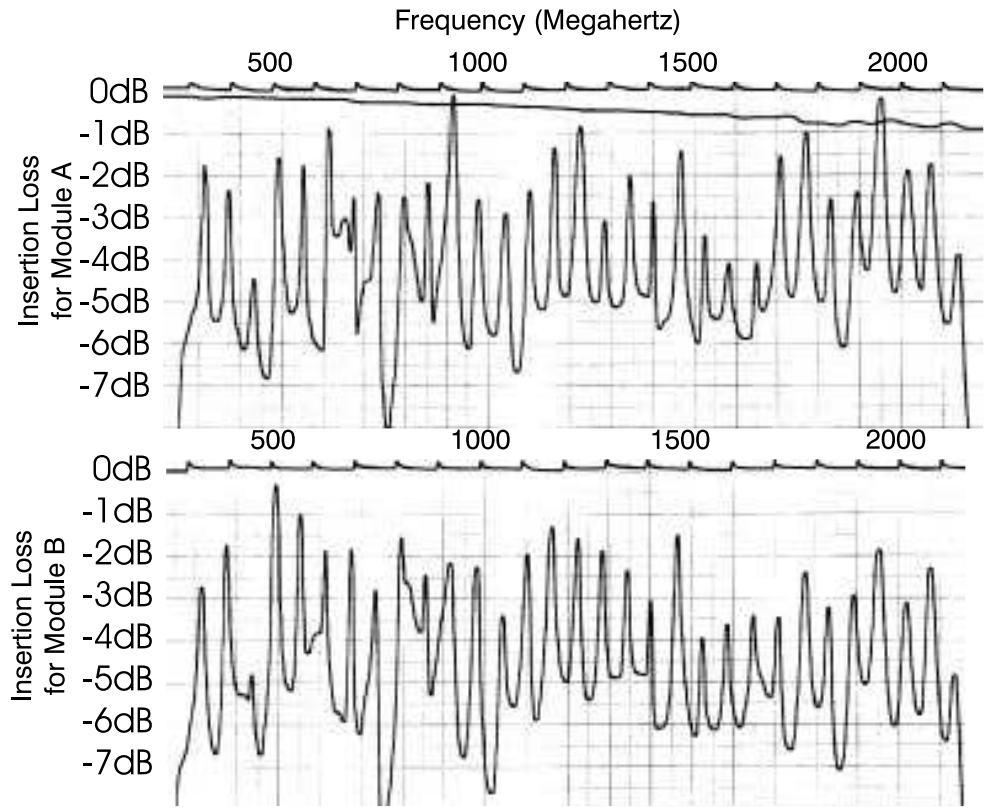}}}
\caption[Insertion loss (gain) of filterbank modules]{
This figure shows the filterbank transmission
(or insertion loss) as a function of frequency for both
filterbank modules.  The two horizontal
lines with tick marks define the frequency scale in steps of 100~MHz.
The nearly straight line with small slope near the top of the figure
is the transmission of the cables connecting to the filterbank
modules.
The top curve (with the large ripple) corresponds to
module A (plus connecting cables)
used in telescope 1, while the bottom curve is the
frequency response of module B used in telescope 2.
Ideally these
spectra should be flat, but improper overlap of neighboring filter
bandpasses causes a $\pm$2.5 dB
bandpass ripple (see \S\ref{section:fb_performance}).
\label{fig:FB_gain}}
\end{center}
\end{figure}

\begin{figure}
\begin{center}
\centerline{\epsfxsize=\columnwidth{\epsfbox{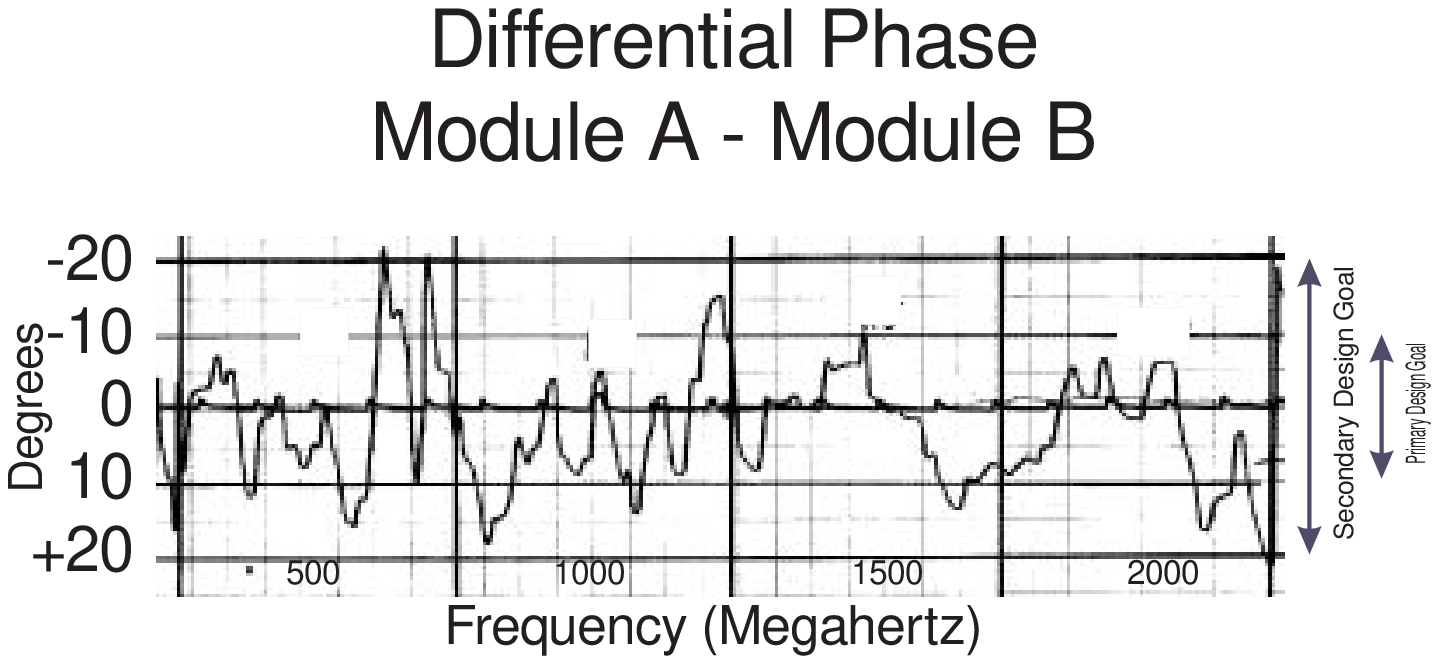}}}
\caption[Relative phase mismatch between filterbank modules]{
This figure shows the relative phase response of the two 
filterbank modules as a function of frequency.
The bandpasses are matched to within 10$\arcdeg$ over 80\% of the
bandwidth, and always within 20$\arcdeg$.
\label{fig:FB_phase}}
\end{center}
\end{figure}

Figure\,\ref{fig:FB_gain} shows the transmission of the two filterbank
modules of the completed filterbank.  An ideal frequency response would
be a flat transmission curve inside the allowed bandpass.  However,
a peak
in the transmission occurs at overlapping band edges due to a design flaw in
the individual filter bandpasses; see Monnier (1999) for further
detail.  The net effect of this mismatch is a slight degradation of
the signal-to-noise ratio.  This occurs because when the RF is finally
detected in a diode, the signal adds up 
coherently as a function of frequency, while the noise adds up incoherently.
In order to accurately estimate the effect on the SNR, 
the phase response of the system must also be characterized.  
A network analyzer has been used to measure the relative phase delay 
for RF signals propagating through the two modules of the filterbank. 
This ``differential phase''
response 
appears in figure\,\ref{fig:FB_phase} and shows that the phase is
matched to $\pm$10$\arcdeg$ over 80\% of the bandpass and is always within
20$\arcdeg$.
Using this information, 
calculation (based on Eq.\,(6.36) in Thompson, Moran, \& Swenson [1986])  
shows that the
filterbank gain ripple only decreases the signal-to-noise ratio by
$\sim$10\% over most of the bandpass.  Further discussion of the
SNR degradation
as a function of frequency can be found in Monnier (1999).

\begin{figure}
\begin{center}
\centerline{\epsfxsize=4in{\epsfbox{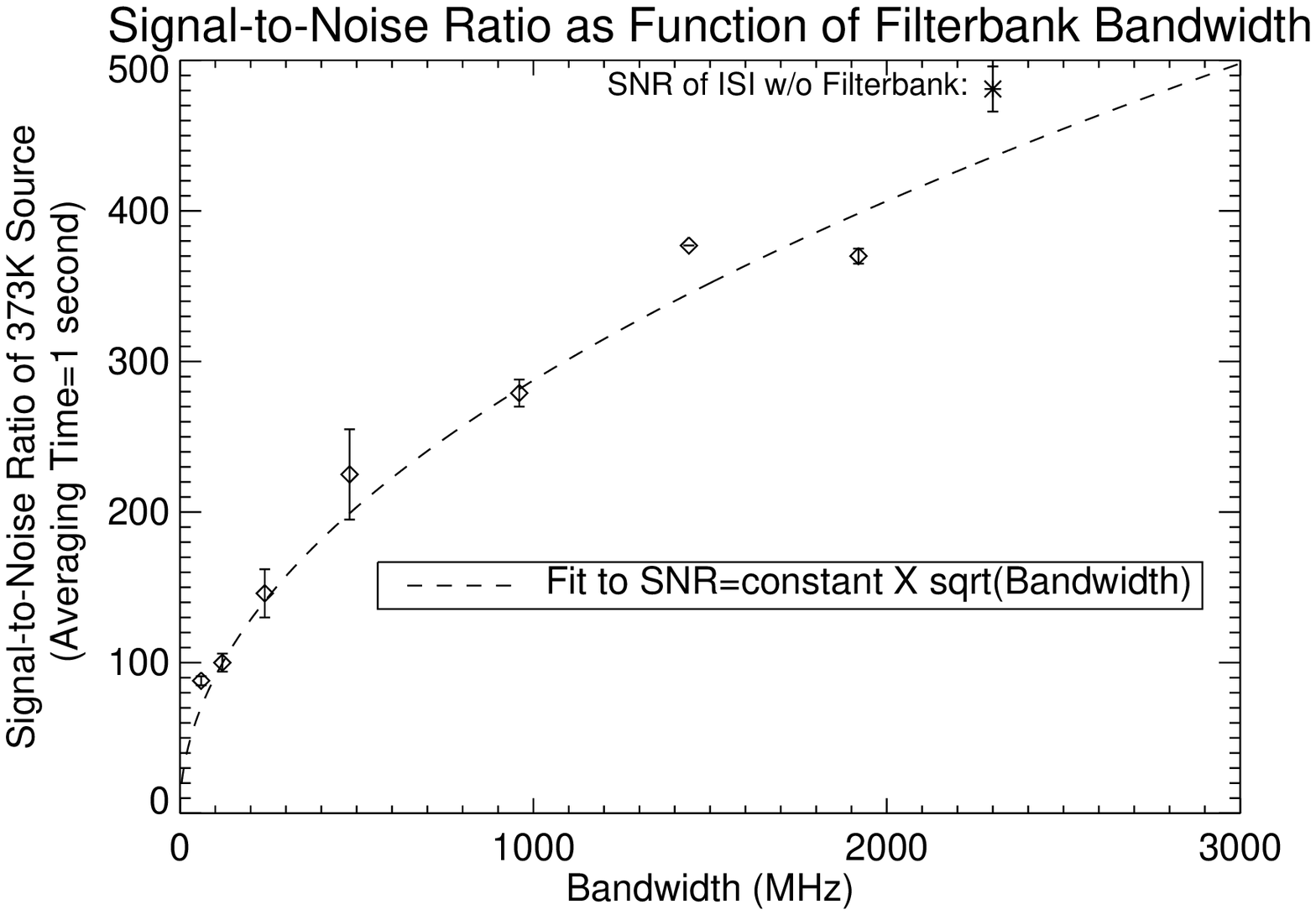}}}
\caption[SNR measurement with filterbank]{A plot of the
signal-to-noise ratio for
observations of a 373K blackbody with and without the filterbank.
This confirms that the system is shot-noise limited and that the
filterbank degrades the sensitivity by about 10\% as expected.
\label{fig:snr}}
\end{center}
\end{figure}

In order to confirm these calculations, signal-to-noise ratio
measurements were made at the telescope while observing a calibration
hotbody source (373\,K).  The results can be found in
figure\,\ref{fig:snr}.  The SNR was measured for a series of bandpass
widths throughout the full frequency range.  There are two major
conclusions that can be drawn from these measurements.  First, the
signal-to-noise ratio has the expected behavior as a function of
bandwidth (SNR$\propto \sqrt{ {\rm Bandwidth}}$).  
%This would not be
%true if excess noise was being generated by filterbank amplifiers after
%filtering, because broadband amplifier noise 
%instead of
%laser shot-noise.  
Secondly, the SNR of the filterbank, when
extrapolated to the effective bandwidth of normal ISI system
($\sim$2300 MHz), is $\sim$10\% low (expected SNR$\sim$500),
confirming the theoretical expectation that the gain ripple causes
only a modest loss in SNR.

\section{Observing Methodology}
\label{section:ir_methodology}
The filterbank observing methodology described in this section
allows a precise and accurate {\em relative} measurement of the
infrared power in two bandpasses.
The first, Bandpass 0 (BP0), is usually only 
a few filters wide (to match the absorption line width)
 while the second, Bandpass 1 (BP1), is typically
much larger (to make an accurate measurement of the continuum).  
The filterbank switches rapidly between these two bandpasses
in order to calibrate signal fluctuations due to seeing and telescope
guiding changes.  This observing strategy makes the ratio of IR power in
BP0 to BP1 insensitive to such sources of fluctuations.

Determining the final calibrated ratio of infrared power in BP0 compared to
BP1 requires 4 nested levels of comparison or chopping (sometimes 5), all 
easily controlled through a combination of hardware and software.
\begin{enumerate}
\item{The sky signal is chopped at 150\,Hz against a cold load to
measure the infrared signal from the sky and minimize the effect of 
low frequency noise in the detection system.}
\item{The filterbank switches between BP0 and BP1 at 2.0833\,Hz in 
order to calibrate out seeing and guiding changes.}
\item{The telescope nods 5$''$ on either side of the star
every 15 seconds to measure 
the changing thermal background of the atmosphere and telescope optics.}
\item{The flat spectrum of the hotbody source (373\,K) is observed
every 5 minutes to measure the overall (frequency-dependent)
gain of the ISI system, precisely 
calibrating the IR power ratios in BP0 to BP1.}
\item{When using lasing transitions of the $^{12}$CO$_2$ molecule, 
a final calibration 
is made for telluric absorption of $^{12}$CO$_2$.  This is done 
be repeating these IR power ratio measurements on the star but using
a nearby 
transition of the CO$_2$ laser local oscillators, one
which has no known molecular line coincidences within
the filterbank's bandpass.  }
\end{enumerate}

IR total power data taken in the above manner are analyzed using an
Interactive Data Language (IDL) reduction package 
which automatically separates out the various
signals and returns a calibrated IR power ratio of BP0 with respect to BP1.

\section{Spectroscopy Results}
As previously emphasized, the lack of multiplexed readouts makes the
acquisition of full-bandwidth, high resolution spectra
very time consuming.  Hence, a series of strategic 
observations were planned to robustly test the experimental
methodology and confirm calibration reliability.  A deep and
broad CO$_2$ absorption line of the Martian atmosphere and a single
NH$_3$ absorption line of IRC\,+10216 were observed for these purposes,
with results described below.

\subsection{Mars}
\label{section:marsline}
Deep absorption lines of Mars have previously been  
measured with a heterodyne
spectrometer using a CO$_2$ laser as a local oscillator by Betz (1977).  
The mid-infrared spectrum of Mars arises from thermal emission of the
warm surface, heated by the Sun.  Since the atmosphere of Mars is composed
largely of $^{12}$CO$_2$, deep absorption lines are formed when viewing the
surface through this molecular blanket.  Mars is roughly as bright
as IRC\,+10216, the brightest stellar mid-infrared source, and its
pronounced spectral features present an ideal test case for calibration and
checks of observing methodology.

\begin{figure}
\begin{center}
\centerline{\epsfxsize=4.5in{\epsfbox{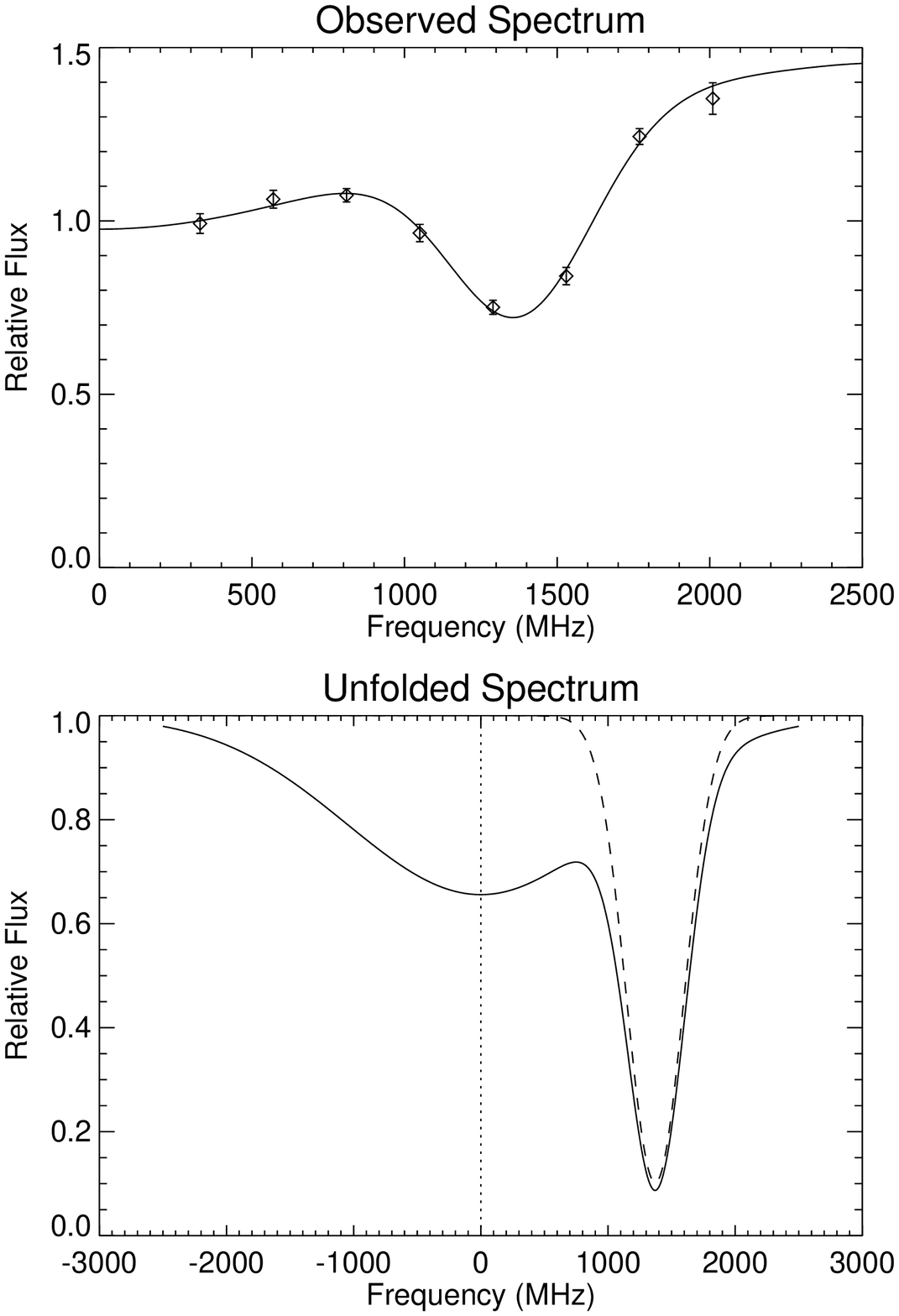}}}
\caption[CO$_2$ line of Mars]{{\em Top Panel:} Mars 
observation of P(16) transition
of $^{12}$CO$_2$ on 1996 Nov 8.  {\em Bottom Panel:} A two-gaussian model 
of the (unfolded) double-sideband spectrum. 
The wide gaussian centered at 0 frequency
comes from telluric absorption, while the narrow, shifted
feature is from the atmosphere of Mars itself.  The relative
velocity of Mars during this observation was -14.54 km/s (1378~MHz
at 10.55\,$\mu$m).
\label{fig:marsline}}
\end{center}
\end{figure}

Mars was observed on 1996 November 08 (UT), with the ISI lasers tuned
to the P(16) transition of $^{12}$C$^{16}$O$_2$ 
at 10.55\,\micronn.  Combinations of 4
consecutive filters ($\Delta\nu=240\,$MHz) were observed across the
entire ISI bandwidth to map out the spectral shape of the 
$^{12}$C$^{16}$O$_2$ P(16)
absorption in the atmospheres of Mars and the Earth.  In each case,
the narrow 4-filter set was compared to the full-bandwidth of the
filterbank by frequency switching at 2.0833~Hz ($\tau$=480~ms) (see
\S\ref{section:ir_methodology}) and integrating for $\sim$5 minutes.
Figure\,\ref{fig:marsline} (top panel)
shows the measured IR power in each of the 4-filter bandpasses 
relative to the average of the entire continuum.  The
presence of the Martian CO$_2$ line
near 1400\,MHz is superimposed
on a shallower, broader absorption feature attributed to the telluric
absorption.  Since the ISI is not able to separate the upper and lower
sidebands, the $\sim$50\% double-sideband 
depth of the Mars absorption line corresponds to
nearly 100\% absorption in a single sideband.  
This ``folded'' spectrum can be decomposed 
into the two sidebands by use of a model for the
spectral shape.  

After decomposition, the double-sideband spectrum in figure~6 (top panel)
is represented by a superposition of
two Gaussian absorption features on a flat continuum.
The center of one Gaussian is fixed at 0\,MHz to fit telluric
absorption, while the other Gaussian center is left free to fit the
location of the Mars absorption core.  The fitting procedure
compensates for the relatively low spectral resolution (240\,MHz) by
smoothing the candidate double-sideband spectrum before folding and
comparing to the data.  The result of this fit shown in the bottom
panel of figure\,\ref{fig:marsline} is in good agreement with 
expectations.  The line center of the Mars line can be predicted based
on the relative velocity of Mars with respect to Earth along the our
line of sight.  NASA ephemerides predict a relative velocity of -14.54
km\,s$^{-1}$ on the date of observation.  The corresponding Doppler
shift is 1378\,MHz at 10.55\,$\micron$, which is good agreement with
the best-fitted center of 1380\,MHz.
%Each data point has an uncertainty of
%$\sim$2\% for a 
%5 minutes observation, consistent with the expected signal-to-noise ratio.

\subsection{IRC +10216: NH$_3$ aQ(2,2)}
The observation of Mars convincingly demonstrated spectral line observations
with the ISI and this filterbank.  However, for observation of 
narrow stellar lines, one
must {\em very}
precisely compensate for the orbital motion of the Earth (doppler
shifts due to the Earth's rotation are slightly less than the
resolution of the filterbank and are neglected).
Software to predict molecular line doppler shifts which 
had previously been used for the
heterodyne measurements of AGB stars in the 1980s (Goldhaber 1988) was
provided by Dr. Betz.

\begin{figure}
\begin{center}
\centerline{\epsfxsize=5in{\epsfbox{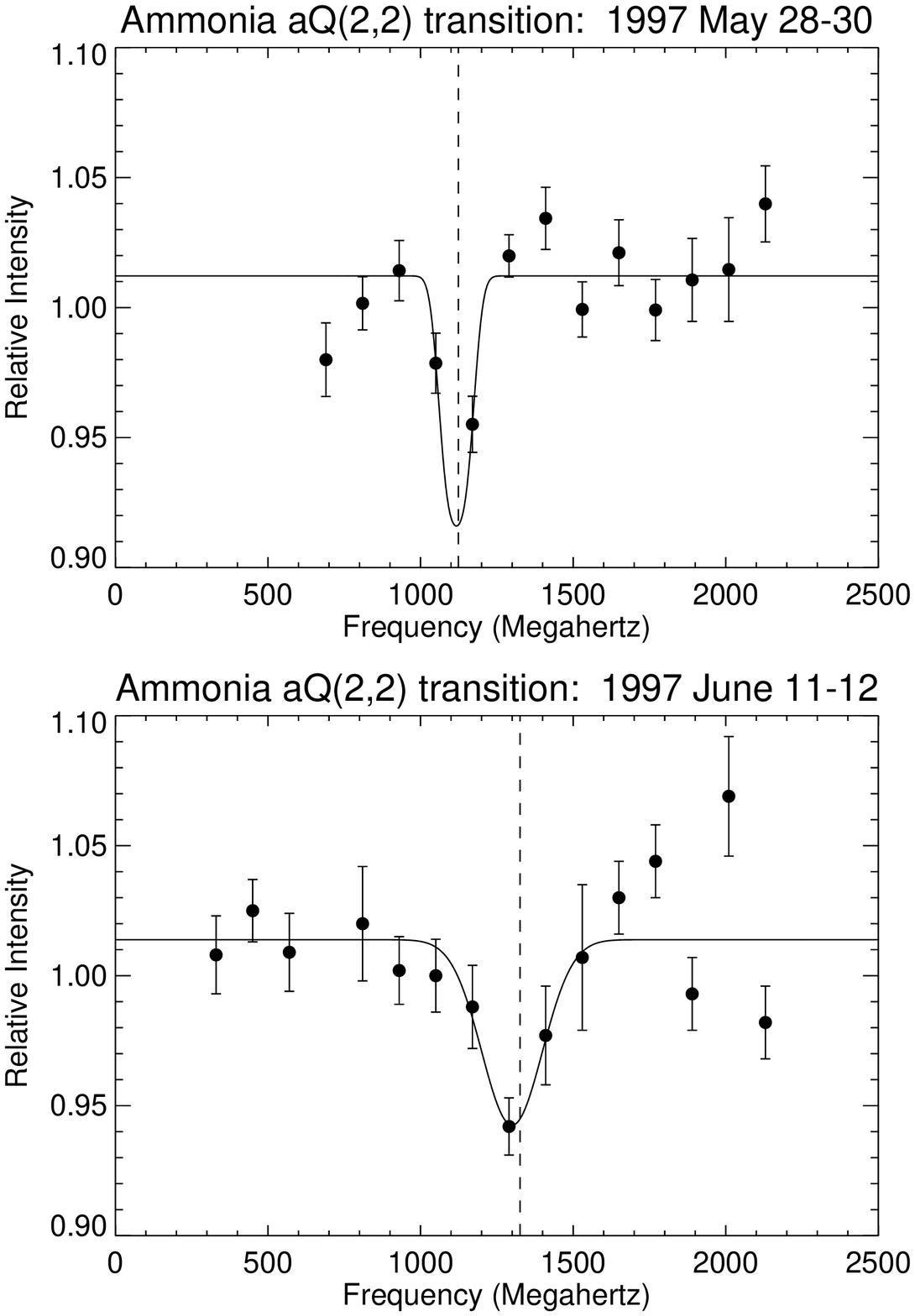}}}

\caption[NH$_3$ line of IRC+10216]{{\em Top Panel:} Observation of aQ(2,2)
transition of NH$_3$ around IRC\,+10216 
on 1997~May~28-30 {\em Bottom Panel:} Observation of
aQ(2,2) transition of NH$_3$ around IRC\,+10216 
on 1997~June~11-12.  Single gaussian fits to the
absorption profiles are shown.  The vertical dashed line in each panel marks
the predicted location of the absorption core, based on previous
observations.  Due to the Earth's orbital motion, there is a frequency shift
between observations which is clearly detected.  
\label{fig:10216_line}}
\end{center} 
\end{figure}

Two test observations of IRC +10216, separated by two weeks, were made in
Spring of 1997.  The observations were separated
in time to allow the Earth to move along its orbit enough for a 
detectable shift
in the line position to occur.  Furthermore, since the chosen line had
been previously observed by Goldhaber (1988),these observations would
test whether the line depth has remained constant over the last
decade.  Both observations employed combinations of two filters which
resulted in a spectral resolution of 120 MHz, just small enough to resolve the
line as it appeared in Goldhaber ([1988] $\sigma\sim$150 MHz).
Figure\,\ref{fig:10216_line} shows the results of these observations
from May and June 1997.

Gaussian absorption features, smoothed to the 120 MHz resolution of
this observation, were fitted to the data and the line center
frequencies determined.  Table\,\ref{table:10216_line} shows the
observed line center locations and the theoretical predictions. 
The agreement is excellent and justifies 
interferometric observations in the cores of previously observed lines
with confidence in the calculations.  The line detection in May was
only marginal, but the movement of the absorption core is clear.
While poorly determined, the apparent line widths and depths are
consistent with the previous measurements of Goldhaber (1988).

\begin{deluxetable}{cccc}
\tablecaption{
Test of software which calculates Doppler shifts of 
celestial spectral lines: 
NH$_3$ aQ(2,2) around IRC +10216
\label{table:10216_line}}
\tablehead{
\colhead{Date}  & \colhead{Observed Absorption Core} & 
\colhead{Predicted Absorption Core} \\
\colhead{(UT)}  & \colhead{Location (MHz)} & \colhead{Location (MHz)} }
\startdata
1997 May 28-29 & 1117$\pm$8 & 1124 \\
1997 June 11-12 & 1299$\pm$35 & 1326 \\
\enddata
\end{deluxetable}

\subsection{Final calibration check}
One further set of tests was performed to determine the accuracy of the IR
power ratio measurements.  The infrared power ratio in
two different filterbank bandpasses, 
denoted by Bandpass 0 and Bandpass 1,
was measured in
Fall 1998 on 8 separate occasions while observing featureless stellar
spectra (i.e, no known line coincidences).  The average IR power of
the full dataset in Bandpass 0 compared to Bandpass 1 was
0.996$\pm$0.005, consistent with unity.  

\section{Interferometry on Spectral Lines: Methodology}
Previous sections discussed the methodology of taking infrared power
measurements on and off of a spectral line.  However to perform
interferometry, the fringe signal must be measured on and off the
line as well.  
%A calibrated visibility measurement requires the fringe power 
%as well as the infrared power to be measured.  
Measurements of the fringe power ratio on and off a spectral line
can be combined with the IR power ratio measurements
to yield the {\em visibility} ratio on and off the line, according to 
the formula (Hale\etal 2000):
\begin{equation}
{\rm Visibility~Ratio} = \sqrt{  \frac{ {\rm Fringe~Power~Ratio} }
{ {\rm IR~Power~Ratio~Telescope~1} \times {\rm IR~Power~Ratio~Telescope~2} } }
\end{equation}
This section will 
detail the methodology to
obtain a well-calibrated fringe power ratio using the ISI filterbank system.

\subsection{Frequency switching (bandpass chopping)}
As with the IR power ratio, for fringe power measurement
the ISI
filterbank is preset so that a narrow bandpass is selected to coincide
with the core of a molecular absorption feature; this
bandpass is referred to as Bandpass 0 (BP0).  A broad bandpass away
from the absorption core, Bandpass 1 (BP1), is selected as a
reference.  BP1 is generally much larger than BP0 so that shot noise
in the BP0 measurement is the dominant noise term in the ratio of BP0
to BP1.  A switching signal, generated and recorded by the 
real-time computer, is used to chop between these two bandpasses.

The frequency switching rate is adjustable and can be slowed 
to take advantage of good
seeing conditions.  In ordinary seeing conditions, a switching period
of 240~ms is used, hence each bandpass is observed for blocks of
120~ms at a time.  Even during poor
atmospheric conditions, this provides for
accurate calibration of fringe
amplitudes and (relative) phase, allowing shot-noise (in BP0) to
dominate the measurement error of the fringe power.  For observing
nights with excellent seeing, a slower frequency switching rate allows longer
coherent integration due to the longer atmospheric coherence time.  In
these cases, a 480~ms switching period was typically used.  

For observations of an absorption line, the local oscillators were
tuned to the appropriate lasing transition to bring the target
molecular transition into the RF bandpass of the ISI detectors.  
Frequency switching was initiated and the correlator output voltage 
recorded (at 500~Hz) for an observation lasting approximately 5
minutes.

\subsection{Delay line gain correction}
As with any interferometer, the ISI uses a variable ``delay line'' to 
compensate for the changing geometric delay of the light coming from 
a source moving across the sky.  
While an automatic gain control circuit has been
implemented in the delay line to keep the average transmission constant
as the 
delay length changes, the spectral shape of the transmission
does not remain precisely constant.  
The frequency-dependent transmission (or
``gain'') of the delay line can change by 5-10\% 
over the course of a few minute
observation, with a direct dependence on the specific length of delay in place.
As can be seen in figure\,\ref{fig:isi}, the
filterbank is located in the RF chain {\em after} the delay line and
hence these changes corrupt the desired measurement of the fringe
amplitude ratio on and off of the spectral line.  On the 4m baseline,
which was used for most of the filterbank observations, the delay line
changed every 20-60 seconds, depending on the source elevation.
The associated spectral changes had to be calibrated in order to achieve
accurate calibration of the fringe amplitude ratio of BP0 to BP1.

A real-time module was written and incorporated into the ISI
delay line software in the control computer for this purpose.  When activated,
this code instructs the data acquisition program to record into a file
the exact time and settings of the delay line during an
observation.  After each $\sim$5-minute measurement, the frequency dependent
gain changes of the delay line were immediately calibrated.  This was
done by observing a known flat continuum source (the ISI hotbody
source at 373\,K) and measuring the IR power in BP0 and BP1 while
cycling through the just-used delay line settings.  While this
method corrects for variations in the gain amplitude of the delay
line, it does not correct possible phase mismatches which, if more
than $\sim$20$\arcdeg$, cause residual miscalibration in the
fringe amplitude in BP0 with respect to BP1.  Calibration of this
effect is discussed in the next section.

\subsection{Correlator drifts/LO switching}
\label{section:line_switch}
Extensive engineering observations during Fall 1997 and Spring 1998
revealed that the spectral response of the fringe measurement, or
correlator circuitry, showed drifts on the time-scale of 0.5 hr,
unrelated to the delay line settings.  Despite significant engineering
efforts, the source of these variations could not be controlled and
one additional calibration procedure was adopted.  Observations of a
stellar source with a featureless spectrum were required for an
absolute calibration of the fringe amplitude ratio on and off of a
spectral line (BP0/BP1).  After each 5-minute observation on the
target spectral region using the appropriate laser LO line, the LOs
were adjusted to a neighboring lasing transition and the process
repeated.  Hence interleaved observations were obtained under
identical conditions except for the LO wavelength (which differed
by $\sim$0.015\micron).  The true ratio of fringe amplitude in BP0 to
BP1 is then the BP0/BP1 ratio at the target LO wavelength, divided by the
BP0/BP1 ratio at the calibration LO wavelength.  This method (``LO
switching''), while involving much off-source integration and repeated
adjustments of the laser grating in order to change the laser
frequency to the alternating transitions, is extremely robust, and has
no known systematic biases.

\subsection{Software for analysis of fringe data}
Custom software written in IDL was used to read the recorded correlator
voltage and bandpass-switching signals.  This software took a power
spectrum of each separate (usually 120~ms) block of fringe data for
BP0 and BP1.  The fringe power was measured around the
fringe frequency (100~Hz) and the
broadband noise signal was subtracted by sampling symmetric frequency
bins $\sim$20~Hz on either side of the main peak.  Careful
measurements during the engineering phase of the filterbank
commissioning established the noise power spectrum after the correlator
circuit to be linear with frequency around 100~Hz to within 0.5\%.  
Data for each delay line setting were then collated and separately
calibrated.  

The mean and standard deviation of the fringe power in each bandpass
and for each delay line setting were then determined by a simple
bootstrap analysis (see Monnier [1999] for more detail).  In addition,
the fringe phase in BP0 can be subtracted from that measured in BP1.
The frequency switching was rapid enough to eliminate atmospheric
variations, essentially ``freezing'' the atmosphere, and the relative
fringe phase in BP0 with respect to BP1 can hence be properly
averaged (in the complex plane), a form of ``phase-referencing.''
After a total of a few nights of observing time on IRC +10216 using a
narrow bandpass of about 180\,MHz ($\Delta\lambda\sim$0.00007\micron),
the visibility amplitude ratio of BP0 to BP1 was measured to a
precision of $\sim$1\% and the relative fringe phase to less than a
few degrees.  The first interferometric results on spectral lines
are presented in Paper III of this series.

\section{Conclusions}
An RF filterbank has been constructed and used with the ISI 
interferometer.
The ISI filterbank functions well as a spectrometer, although the data
rate is low.  The filterbank system was not designed to be an efficient
spectrometer, but rather to be an effective first-generation tool to
harness the spectral resolving power of the ISI for interferometry.  
The observations
presented here confirm the ability of the ISI to do such
work, and an observing methodology for combining interferometric observing
with filterbank observation has been outlined.
This methodology has been used successfully for observing fringe visibilities
on and off of a number of molecular absorption lines.  Preliminary results
are discussed in Monnier (1999), and a more complete analysis follows in
Papers II and III of this series.

This new capability of the ISI has so far been used to measure the
location of molecular formation zones for NH$_3$ and SiH$_4$ around
the carbon star IRC\,+10216 and NH$_3$ around the red supergiant VY~CMa.  However,
strong spectral absorption features of these same molecules could also
be investigated for IRC\,+10420, $o$~Cet, and NML Cyg, whose line
depths and shapes have already been measured (\cite{mclaren80};
\cite{betz85}; \cite{goldhaber88}). Measurements of angular diameters 
of red supergiants and Miras on and off their recently discovered
mid-infrared water lines (\cite{jennings98}) is also potentially
rewarding.  These lines are thought to form in the stellar photosphere
and may affect the interpretation of previous angular diameter
measurements.  The use of high spatial and spectral resolution line
observations in between water lines could potentially allow the ``true''
continuum diameters to be measured, uncorrupted by line blanketing.
Lastly, fine structure and recombination lines around some
emission-line stars (e.g., MWC\,349,
\cite{quir97}) can be observed using the CO$_2$ local oscillators, and
could make unique measurements of the emitting regions.

While the potential for interesting new results is rich, the
non-multiplexed output of this particular filterbank system makes it
very difficult to observe new sources in detail when the spectral
line profiles have not been previously measured by a multiplexing
spectrometer.  The lines around some sources ($o$ Ceti, for instance)
have been observed to change in time, and new spectra are essential
for interpreting additional interferometric line data.

\acknowledgements
{ We want to recognize the computer programming of Manfred Bester and
Carl Lionberger which was essential for the filterbank to
interface with the ISI control system.
This work is a part of a
long-standing interferometry program at U.C. Berkeley, supported by
the National Science Foundation (Grant AST-9221105,
AST-9321289, and AST-9731625)
and by the Office of Naval Research (OCNR N00014-89-J-1583).}

%\figcaption{
%This figure shows 
%\label{fig:blah}}


\begin{thebibliography}{} 

\bibitem[Bester et al. 1996]{bester96} Bester, M., Danchi, W. 
C., Hale, D., Townes, C. H., Degiacomi, C. G., Mekarnia, D. \& Geballe, T. 
R. 1996, \apj, 463, 336 

\bibitem[Betz et al. 1976]{betz76} Betz, A. L., 
Johnson, M. A., McLaren, R. A. \& Sutton, E. C. 1976, \apjl, 208, L141 
 
\bibitem[Betz 1977]{betz77} Betz, A. L. 1977, 
University of California at Berkeley, PhD Dissertation

\bibitem[Betz \& Goldhaber 1985]{betz85} Betz, A. L. \& 
Goldhaber, D. M. 1985, in Mass Loss from Red Giants,
ed. M. Morris \& B. Zuckerman (Dordrecht: Reidel), 83

%\bibitem[Bieging \& Tafalla 1993]{bieging93} Bieging, John H. \& 
%Tafalla, Mario 1993, \aj, 105, 576 

%\bibitem[Boyle et al. 1994]{boyle94} Boyle, R. J., Keady, J. 
%J., Jennings, D. E., Hirsch, K. L. \& Wiedemann, G. R. 1994, \apj, 420, 863 

\bibitem[Danchi et al. 1994]{danchi94} Danchi, W. C., Bester, 
M., Degiacomi, C. G., Greenhill, L. J. \& Townes, C. H. 1994, \aj, 107, 1469 

%\bibitem[Efron \& Tibshirani 1993]{bootstrap} Efron, B., \& Tibshirani, R. J.
%1993, {\em An introduction to the bootstrap}, Chapman and Hall: New York

%\bibitem[Glassgold \& Mamon 1992]{glassgold92} Glassgold, A.E. \& 
%Mamon, G.A. 1992, Chemistry and Spectroscopy of Interstellar Molecules, 261 

%\bibitem[Glassgold 1998]{glassgold98} Glassgold, A. E. 
%1998, IAU Symposia, 191, 337

%\bibitem[Goldhaber \& Betz 1984]{goldhaber84} Goldhaber, D. M. \& 
%Betz, A. L. 1984, \apjl, 279, L55 

\bibitem[Goldhaber 1988]{goldhaber88} Goldhaber, D. M. 1988,
University of California at Berkeley, PhD Dissertation

%\bibitem[Gray et al. 1977]{gray77}
% Gray, D. L., Robiette, A. G., \& Johns, J. W, 1977,
%Molecular Physics, 34, 1437

%\bibitem[Groenewegen 1997]{groenewegen97} Groenewegen, M.A.T. 1997, 
%\aap, 317, 503 

\bibitem[Hale et al. 2000]{hale2000} Hale, D. D. S., et al.
2000, \apj, Submitted

%\bibitem[Haniff \& Buscher 1998]{haniff98} Haniff, C. A. \& 
%Buscher, D. F. 1998, \aap, 334, L5 

%\bibitem[Herbig \& Zappala 1970]{herbig70} Herbig, G. H. \& 
%Zappala, R. R. 1970, \apjl, 162, L15 

%\bibitem[Herbig 1972]{herbig72} Herbig, G. H. 1972, \apj, 172, 
%375 

\bibitem[Holler 1999]{holler99} Holler, C. 1999, 
Ludwig-Maximilians-Universitaet,
Master's Thesis

\bibitem[Isaak, Harris \& Zmuidzinas 1999]{isaak99} Isaak, K., 
Harris, A. I. \& Zmuidzinas, J. 1999, in Highly Redshifted Radio Lines, ASP 
Conf. Series Vol. 156, Ed. by C. L. Carilli, S. J. E. Radford, K. M. 
Menten, \& G. I. Langston, p. 86

%\bibitem[Ivezic \& Elitzur 1996]{ivezic96} Ivezic, Z. \& 
%Elitzur, M. 1996, \mnras, 279, 1019 

\bibitem[Jennings \& Sada 1998]{jennings98} Jennings, Donald E. \& 
Sada, Pedro V. 1998, Science, 279, 844 

%\bibitem[Kastner \& Weintraub 1998]{kw98} Kastner, J. H. \& 
%Weintraub, D. A. 1998, \aj, 115, 1592 

%\bibitem[Keady 1982]{keady82} Keady, J. J.  1982,
%New Mexico State University, Ph.D. Dissertation

%\bibitem[Keady, Hall \& Ridgway 1988]{keady88} Keady, J. J., 
%Hall, D. N. B. \& Ridgway, S. T. 1988, \apj, 326, 832 

%\bibitem[Keady \& Ridgway 1993]{keady93} Keady, J. J. \& 
%Ridgway, S. T. 1993, \apj, 406, 199 

%\bibitem[Knapp et al. 1982]{knapp82} Knapp, G. R., Phillips, T. 
%G., Leighton, R. B., Lo, K. Y., Wannier, P. G., Wootten, H. A. \& Huggins, 
%P. J. 1982, \apj, 252, 616 

%\bibitem[Lada \& Reid 1978]{ladareid78} Lada, C. J. \& Reid, M. J. 
%1978, \apj, 219, 95 

%\bibitem[Leitch-Devlin \& Williams 1985]{sticky85} 
%Leitch-Devlin, M. A. \& Williams, D. A. 1985, \mnras, 213, 295 

\bibitem[Lipman 1998]{lipman98} Lipman, E. A. 1998, University of
California at Berkeley, PhD Dissertation

%\bibitem[Maihara et al. 1976]{maihara76} Maihara, T., Noguchi, 
%K., Oishi, M., Okuda, H. \& Sato, S. 1976, \nat, 259, 465 

%\bibitem[Marvel 1996]{marvel96} Marvel, K. B  1996,
%New Mexico State University, Ph.D. Dissertation

%\bibitem[Mathis, Rumpl \& Nordsieck 1977]{mrn} Mathis, J. 
%S., Rumpl, W. \& Nordsieck, K. H. 1977, \apj, 217, 425 

%\bibitem[McCarthy 1979]{mccarthy79} McCarthy, D. W., Jr. 1979, IAU 
%Colloq. 50: High Angular Resolution Stellar Interferometry, 18
 
%\bibitem[McCarthy, Howell \& Low 1980]{mccarthy80} McCarthy, D. W., 
%Howell, R. \& Low, F. J. 1980, \apjl, 235, L27 

\bibitem[McLaren \& Betz 1980]{mclaren80} McLaren, R. A. \& Betz, 
A. L. 1980, \apjl, 240, L159 

%\bibitem[Mihalas Kunasz \& Hummer 1975]{mihalas75} Mihalas, D., 
%Kunasz, P. B. \& Hummer, D. G. 1975, \apj, 202, 465 

%\bibitem[Monnier et al. 1997]{jdm:nmlcyg} Monnier, J. D., et al. 
%1997, \apj, 481, 420 

\bibitem[Monnier 1999]{mythesis} Monnier, J. D. 1999, University
of California at Berkeley, PhD Dissertation

%\bibitem[Monnier, Geballe, \& Danchi 1998]{mgd98} Monnier, J. 
%D., Geballe, T. R. \& Danchi, W. C. 1998, \apj, 502, 833 

%\bibitem[Monnier et al. 1999a]{monnier99a} Monnier, J. D., Tuthill, 
%P. G., Lopez, B., Cruzalebes, P., Danchi, W. C. \& Haniff, C. A. 1999a, 
%\apj, 512, 351 

%\bibitem[Monnier, Geballe \& Danchi 1999]{mgd99} Monnier, J. 
%D., Geballe, T. R. \& Danchi, W. C. 1999, \apj, 521, 261 

%\bibitem[Nakanaga et al. 1985]{nakanaga85}
%Nakanaga, T., Kondo, S. \& Saeki, S. 1985, 
%Journal of Molecular Spectroscopy, 112, 39

%\bibitem[Neugebauer \& Leighton 1969]{irc69} Neugebauer, G. 
%\& Leighton, R. B. 1969, {\em Two micron sky survey : 
%A preliminary Catalogue}
%(NASA SP-3047 [Washington, D.C.: Government Printing Office])

%\bibitem[Ossenkopf Henning \& Mathis 1992]{ohm92} Ossenkopf, 
%V., Henning, Th. \& Mathis, J. S. 1992, \aap, 261, 567 

\bibitem[Quirrenbach et al. 1997]{quir97} Quirrenbach, A., 
Thum, C., Martin-Pintado, J. \& 
Matthews, H. E. 1997, \baas, 191, 4719 

%\bibitem[Richards Yates \& Cohen 1998]{richards98} Richards, A. M. 
%S., Yates, J. A. \& Cohen, R. J. 1998, \mnras, 299, 319 

%\bibitem[Rouleau \& Martin 1991]{rm91} Rouleau, Francois \& 
%Martin, P. G. 1991, \apj, 377, 526 

%\bibitem[Shu 1991]{shu91}
%Shu, F. H. {\em The physics of astrophysics. Vol.1: Radiation},
%University Science Books (Mill Valley, CA), 1991

%\bibitem[Le Sidaner \& Le Bertre 1996]{sidaner96} Le Sidaner, P. 
%\& Le Bertre, T. 1996, \aap, 314, 896 

%\bibitem[Skinner, Meixner \& Bobrowsky 1998]{skinner98} Skinner, 
%C. J., Meixner, M. \& Bobrowsky, M. 1998, \mnras, 300, L29 

%\bibitem[Smith et al. 1999]{smith99} Smith, N., Humphreys, R. 
%M., Krautter, J., Gehrz, R. D., Davidson, K., Jones, T. J. \& Hubrig, S. 
%1999, American Astronomical Society Meeting, 194, 1306 

\bibitem[Thompson, Moran \& Swenson 1986]{thompson86} Thompson, A. 
Richard, Moran, James M. \& Swenson, George W. 1986, 
{\em Interferometry and synthesis in radio astronomy}, 
Wiley-Interscience: New York

%\bibitem[Toombs et al. 1972]{toombs72} Toombs, R. I., Becklin, 
%E. E., Frogel, J. A., Law, S. K., Porter, F. C. \& Westphal, J. A. 1972, 
%\apjl, 173, L71 

%\bibitem[Townes \& Shawlow 1975]{townes75} Townes, C. H., \& Schawlow, A. L.
%1975, {\em Microware Spectroscopy}, Dover Publications, Inc. : New York

%\bibitem[Tuthill, Monnier \& Danchi 1998]{tuthill98b} Tuthill, P. G.,
%Monnier, J. D., \& Danchi, W. C. 1998, IAU Symposia, 191, 331

%\bibitem[Urban et al. 1983]{urban83} 
%Urban, S. , Papousek, D., Kauppinen, J., Yamada, K., \& Winnewisser, G.
%1983, Journal of Molecular Spectroscopy, 101, 1

%\bibitem[Weigelt et al. 1998]{weigelt98a} Weigelt, G., Balega, Y., 
%Bloecker, T., Fleischer, A. J., Osterbart, R. \& Winters, J. M. 1998, \aap, 
%333, L51 

%\bibitem[Winters et al. 1998]{winters98} Winters, J. M., Keady, 
%J. J., Fleischer, A. J.
%\& Gauger, A. 1998, A Half Century of Stellar Pulsation Interpretation: 
%A Tribute to Arthur N. Cox, edited by Paul A. Bradley and Joyce A. Guzik, 
%Proceedings of a Conference held in Los Alamos, NM 16-20 June 1997, ASP 
%Conference Series \#135, p. 337., 337 

%\bibitem[Winters et al. 1995]{winters95} 
%Winters, J. M., Fleischer, A. J., Gauger, A. \& Sedlmayr, E. 1995, \aap, 
%302, 483 

%\bibitem[Worley 1972]{worley72} Worley, Charles E. 1972, \apjl, 
%175, L93 

\end{thebibliography}
\end{document}